\documentclass[a4paper]{article}

\usepackage[pages=all, color=black, position={current page.south}, placement=bottom, scale=1, opacity=1, vshift=5mm]{background}

\usepackage[margin=1in]{geometry} 
\usepackage{graphicx}
\usepackage{subcaption}
\usepackage{amsmath}
\usepackage{amsthm}
\usepackage{amssymb}
\usepackage{authblk}

\usepackage[utf8]{inputenc}
\usepackage{hyperref}
\hypersetup{
	unicode,
	pdfauthor={Author One, Author Two, Author Three},
	pdftitle={A simple article template},
	pdfsubject={A simple article template},
	pdfkeywords={article, template, simple},
	pdfproducer={LaTeX},
	pdfcreator={pdflatex}
}

\usepackage[numbers]{natbib}

\theoremstyle{plain}

\theoremstyle{definition}

\usepackage{graphicx, color}
\graphicspath{{fig/}}

\usepackage{mathtools}

\usepackage{algorithm, algpseudocode} 
\usepackage{mathrsfs} 

\usepackage{lipsum}

\title{Towards mechanistic understanding in a data-driven weather model: internal activations reveal interpretable physical features}

\author[1]{Theodore MacMillan \thanks{tmacmill@stanford.edu}}
\author[1]{Nicholas T. Ouellette \thanks{nto@stanford.edu}}

\affil[1]{Stanford University}
\date{}
\begin{document}
\maketitle

\begin{abstract}

Large data-driven physics models like DeepMind’s weather model GraphCast have empirically succeeded in parameterizing time operators for complex dynamical systems with an accuracy reaching or in some cases exceeding that of traditional physics-based solvers. Unfortunately, how these data-driven models perform computations is largely unknown and whether their internal representations are interpretable or physically consistent is an open question. Here, we adapt tools from interpretability research in Large Language Models to analyze intermediate computational layers in GraphCast, leveraging sparse autoencoders to discover interpretable features in the neuron space of the model. We uncover distinct features on a wide range of length and time scales that correspond to tropical cyclones, atmospheric rivers, diurnal and seasonal behavior, large-scale precipitation patterns, specific geographical coding, and sea-ice extent, among others. We further demonstrate how the precise abstraction of these features can be probed via interventions on the prediction steps of the model. As a case study, we sparsely modify a feature corresponding to tropical cyclones in GraphCast and observe interpretable and physically consistent modifications to evolving hurricanes. Such methods offer a window into the black-box behavior of data-driven physics models and are a step towards realizing their potential as trustworthy predictors and scientifically valuable tools for discovery.

\end{abstract}

\section{Introduction}

Data-driven physics models have achieved state-of-the-art performance in predicting the time evolution of dynamical systems \cite{Lai2025Panda:Dynamics}, especially in the context of high-dimensional, data-rich domains like atmospheric forecasting \cite{Price2025ProbabilisticLearning, Bi2022Pangu-Weather:Forecast}, and operate at a fraction of the computational cost of traditional physics-based solvers \cite{Lam2023LearningForecasting}. However, they are largely opaque and provide few guarantees of adherence to known laws of physics, issues that pose a significant trust barrier to their wide-scale adoption \cite{Thorpe2025MixedForecasting}. This concern stems from two major open questions about the data-driven approach: whether models actually encode laws of physics, and (relatedly) whether they can reliably generalize beyond their training data. In the context of data-driven weather models, for example, it has been argued that models can fail to generalize in a way that results in an inability to predict extremes \cite{Zhang2025NumericalExtremes, Sun2025CanCyclones}. It is therefore essential to ask what patterns and abstractions these models actually encode, and whether these abstractions correspond to the interpretable abstractions of physics. The scientific value of---and empirical trust in---data-driven model predictions will likely follow.

Related interpretability questions have also arisen in the development and deployment of large language models (LLMs), where although strict adherence to physical laws is not a concern, model transparency holds a significant premium. It is largely in this context that the nascent field of \textit{mechanistic interpretability} \cite{Bereska2024MechanisticReview} has been developed, a set of tools that seeks to provide an account for the internal representations held in large data-driven models. Researchers have successfully extracted interpretable features in intermediate processing layers of language models \cite{bricken2023monosemanticity}, discovered circuits of these features that causally influence one another to direct model output \cite{Dunefsky2024TranscodersCircuits, lindsey2025biology}, and demonstrated the ability to steer model behavior by precisely manipulating these internal concepts \cite{templeton2024scaling}. Recent work has extended this research to protein language models (PLMs) \cite{Simon2025InterPLM:Autoencoders, Gujral2025SparseRepresentations, Adams2025FromModels}, where it has been shown that PLMs learn and leverage biologically relevant concepts.

Here, we explore whether such interpretable abstractions are present in a data-driven physics model. We specifically consider GraphCast \cite{Lam2023LearningForecasting}, a state-of-the-art 36.7M parameter model for weather forecasting trained on 40 years of ERA5 reanalysis data \cite{Hersbach2020TheReanalysis}. Using an unsupervised dictionary learning technique known as a sparse autoencoder (SAE) on the hidden layers of GraphCast, we uncover specific combinations of neurons that encode interpretable abstractions corresponding to well-studied weather patterns, including large-scale precipitation, specific geographic coding, and evident diurnal functions. Among these are annually periodic features that correspond to Arctic and Antarctic ice extent, physical features that dynamically affect the atmosphere but are not present in GraphCast's input or output, showing the model's ability to learn physical representations outside of its training set. We also demonstrate the existence of grid-locked features, spurious and potentially undesirable features activating on the grid representation of GraphCast and often unrelated to underlying weather patterns, illustrating the interplay between interpretability and model development. To probe model representation of specific phenomena, we train logistic probes that map single features onto existing datasets for extreme weather and uncover features encoding tropical cyclones (TCs) and atmospheric rivers (ARs). Finally, we demonstrate a method for probing the physical consistency of these model abstractions, showing that selectively amplifying or attenuating the internal abstraction of a TC leads to stronger or weaker predicted storm evolution, and, via a conservation law and force balance analysis, that such modifications lead to dynamically consistent outputs. Code for the analysis is available at \url{https://github.com/theodoremacmillan/graphcast-interpretability}.

\section{Unsupervised discovery of features}

At each layer of message passing in GraphCast, the nodes and edges update their embeddings via message-passing mediated by multi-layer perceptrons (see Appendix \ref{sec:graphcast} for details). For a given node $i$, one can consider its ``neuron'' activations to be the values of the embedding vector $\mathbf{v}_i\in \mathbb{R}^{n_d}$ at that layer. Early interpretability research in similar embedding-based models treated these neuron activations as the fundamental unit of analysis \cite{Gurnee2023FindingProbing}, finding specific neurons that activated in the presence of interpretable concepts \cite{goh2021multimodal}. However, this mode of analysis was complicated by the presence of ``polysemantic'' neurons---neurons that activate in the presence of many different concepts---and much interpretability research has instead shifted towards studying combinations of neurons as the fundamental unit of analysis \cite{bricken2023monosemanticity}.

\subsection{Sparse autoencoders}

The goal of a sparse autoencoder (SAE) is to learn a set of dictionary vectors corresponding to such interpretable groups of neurons in an unsupervised fashion. In our context, this task amounts to finding $n_l$ feature vectors $\textbf{w}_{j}\in \mathbb{R}^{n_d}, \:  j=1,...,n_l$, such that node embeddings can be reconstructed as sparse linear combinations

\begin{equation}
    \mathbf{v}_i\approx W \boldsymbol{\alpha}_i + \mathbf{b}, \: \: \: ||\boldsymbol{\alpha}_i||_0\leq k ,
\end{equation}
where $W\in \mathbb{R}^{n_d\times n_l}$ is a feature matrix and $\boldsymbol{\alpha}_i$ is a vector of sparse feature activations at a particular node, $\textbf{b}$ is some constant offset, and $k\ll n_d$ is the sparsity parameter. That such a reconstruction is even possible is the subject of the \textit{linear representation hypothesis} \cite{Park2024TheModels}, and we note intriguing connections to traditional data-driven approaches for dynamical systems (discussed further in Appendix \ref{sec:sindy}).

\begin{figure}[ht]
    \centering
    \includegraphics[width=0.95\linewidth,  trim={0cm 0 0cm 0}, clip]{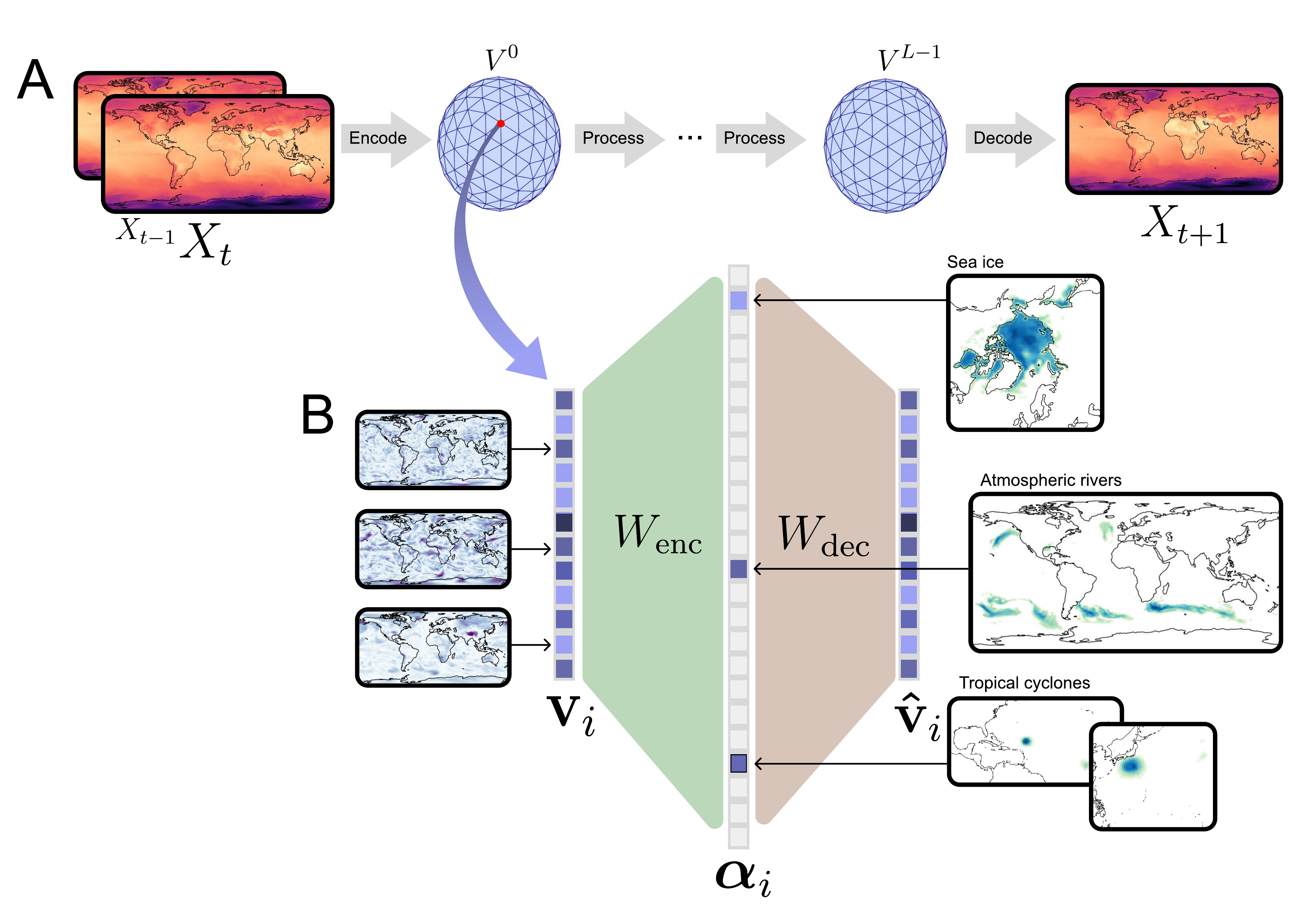}
    \caption{GraphCast interpretability pipeline. \textbf{(A)} Standard encode-process-decode architecture of GraphCast. Atmospheric variables are encoded onto an internal Graph Neural Network (GNN) where processing occurs. \textbf{(B)} Capturing the activations at some intermediate layer of the model, we display dense, uninterpretable nodal embeddings as global maps. By learning a transformation such that these fields can be written as a sparse linear combination of feature vectors, we uncover interpretable abstractions in intermediate GraphCast processing layers.} 
    \label{fig:pipeline}
\end{figure}

SAEs learn this dictionary of feature vectors as well as their activations on given embeddings. For $k$-sparse autoencoders \cite{Gao2024ScalingAutoencoders}, this transformation is represented as

\begin{align}
\boldsymbol{\alpha}_i &= \mathrm{TopK}\!\big(W_{\mathrm{enc}}(\mathbf{v}_i - \mathbf{b}\big)\big) \\
\mathbf{\hat{v}}_i &= W_{dec}\boldsymbol{\alpha}_i + \mathbf{b}
\end{align}
where $W_{enc}\in \mathbb{R}^{n_l\times n_d}$, $W_{dec}\in \mathbb{R}^{n_d\times n_l}$, and $\boldsymbol{\alpha}_i\in \mathbb{R}^{n_l}$, with $n_l \gg n_d$ typically. We therefore seek to represent the dense $n_d$ node embeddings in a much larger $n_l$ sized vector space (the latent space) but with considerable sparsity $k$. Figure \ref{fig:pipeline} illustrates this approach applied to the node embeddings of GraphCast. To encourage informative latents, where magnitude implies per-instance activation, the columns of $W_{dec}$ are constrained to have unit norm. The TopK activation function zeros out activations that are not in the largest $k$ of the instance, directly enforcing the $k$-sparsity constraint.

\subsection{Training}

To train our SAEs, we ran the GraphCast 0.25 degree resolution model with 37 pressure levels for a single prediction step on 40 years of ERA5 snapshots every six hours from 1979 to 2019. We installed hooks in the internal activations, extracting node embeddings $\mathbf{v}_i^l, \: i=1,...,n_n$ at each layer $l=1,...,16$ for each snapshot. GraphCast's largest model contains 40,962 nodes at each message-passing layer, resulting in 2.39 billion node embedding vectors at each layer over our entire dataset. Focusing on an intermediate layer $l=8$, we trained several SAEs across a range of hyperparameters and observed a strong tradeoff between $L_0$ norm (that is, the value of $k$) and reconstruction loss. Details on hyperparameters and training can be found in Appendix \ref{sec:sae}, as well as reported performance on the Pareto-front of sparsity versus reconstruction error. 

\subsection{Feature interpretability}

Performance as measured by these general metrics is a helpful way of comparing architectures against one another or tuning hyperparameters, but does not directly assess the interpretability of the structures we discover from the sparse encoding. The physics-based context of our exploration adds to this difficulty: the notion of interpretability or conceptual coherence is more straightforward in a language context, as concepts in language are described in language. In a dynamical system, what constitutes a `feature' or coherent `concept' is itself a subject of considerable debate. For example, definitions may be based on transport coherence \cite{Haller2015LagrangianStructures, Hadjighasem2017ADetection}, notions of information compressibility \cite{MacMillan2021}, or predictability \cite{Fang2019LocalFlow}. It is difficult to settle on a single, general definition, and correctness ultimately depends as much on context as any particular definition.



\begin{figure}[ht]
    \centering
    \includegraphics[width=1.0\linewidth,  trim={0cm 0cm 0cm 0cm
    }, clip]{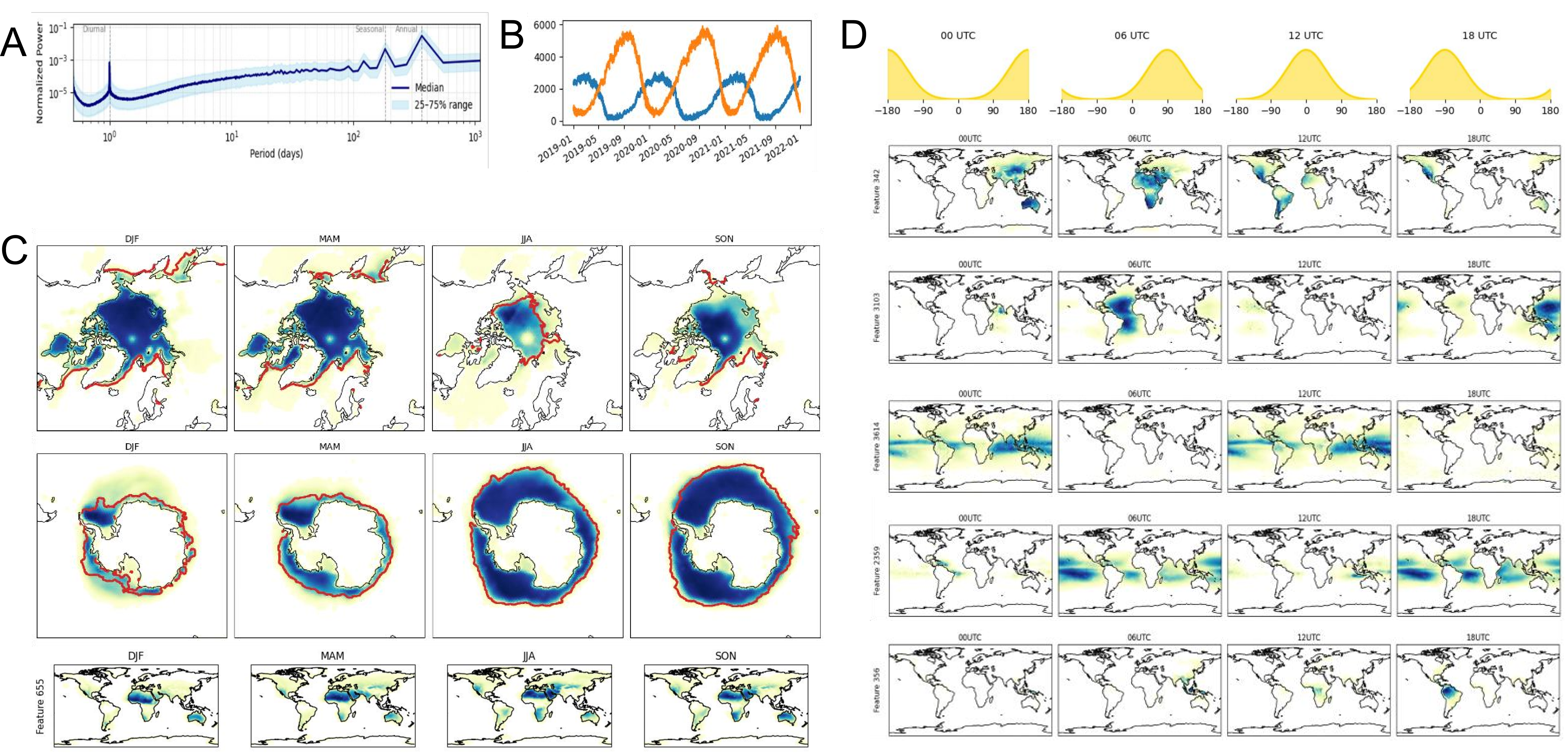}
    \caption{Features on many timescales. \textbf{(A)} Power spectrum of global mean activation over time. Distinctive peaks indicate the existence of features with diurnal, seasonal, and annual oscillations. \textbf{(B)} Two example seasonal feature time series, one activating in the northern hemisphere winter and the other in the southern hemisphere winter. \textbf{(C)} Feature 1710 shows strong seasonality and its activation tracks northern sea-ice extent (overlaid in red), even though no ice extent information is processed by GraphCast. Feature 1437 shows similar strong seasonality but instead tracks southern sea-ice extent (overlaid in red). Another annual feature, feature 655, corresponds to surface heating in desert regions and seasonally migrates. \textbf{(D)} Some example strongly diurnal features. From top to bottom: daytime activation in especially arid regions; ocean basin activation in early morning; precipitation patterns resembling the ITCZ; corresponding negative of precipitation patterns, or especially dry ocean regions; rain forests, activating primarily in the Amazon during the day but also strongly in Indonesia and Africa during their respective sunlight hours.} 
    \label{fig:timescales}
\end{figure}

These difficulties aside, the context of our model allows us to apply some physical intuition to extract interpretable features. We might expect, for instance, that some features are likely periodic on time scales that correspond to seasonal and daily patterns in atmospheric dynamics. In Figure \ref{fig:timescales}A, we show the averaged power spectrum of all learned features, finding distinctive peaks at half-daily, daily, half-yearly, and yearly periods. Diurnal features roughly fall into the first two of these categories, while seasonal features fall into the latter two. The reason for the multiple distinctive peaks is that although diurnal features are active on a daily period for a given part of the globe, they also activate during the day on the other side of the planet, leading to a twice-daily overall signature. We observe a similar phenomenon for seasonal features, where a phenomenon like the polar wind may activate strongly in the northern hemisphere winter, and then again strongly in the southern hemisphere winter, resulting in a twice-yearly period.

We then looked more closely at features where the energy was strongly concentrated in the diurnal or seasonal band. At the seasonal timescale, we discovered two interesting features that respectively appear to track arctic and antarctic sea-ice extent throughout the year. They are shown in Figure \ref{fig:timescales}C along with reference sea-ice extents taken from the Sea Ice Index, V4 dataset provided by the National Snow and Ice Data Center \cite{Windnagel2025NationalAnalysis}. It is particularly notable that these features exist in the model, because ice extent is not an explicit variable processed by GraphCast; thus, we venture that it must be dynamically inferred. This finding suggests that not only are deep-learning weather models able to infer missing information about the state of the globe in order to make dynamical predictions, but also that such information is extractable with completely unsupervised methods (in our case, an SAE). A similar idea was hinted at in \cite{Kochkov2024NeuralClimate}, where the learned physical tendencies of the data-driven complement to a physics-driven dynamical core showed some signature of interpretability. We note, however, that although the sea-ice features tend to align with the ice extent, we demonstrate only correlation here and a more rigorous causal analysis would be necessary for a more complete account. Also shown in Figure \ref{fig:timescales}C is a more typical seasonal feature (feature 655), which corresponds to what appears to be a seasonally varying heating pattern.


Observing large-magnitude activating features with primarily diurnal variation, we find a mix of interpretable and less interpretable structures. Figure \ref{fig:timescales}D displays some interesting cases. Feature 342 activates in the early morning in especially arid regions, possibly indicating a dynamical connection to surface heating. Feature 3103 has a similar correlation, but only in ocean basins. Features 3614 and 2359 correspond to patterns of high and low rainfall anomaly, the former closely tracking the regions known as the intertropical convergence zone (ITCZ) \cite{Liu2020Observed1998-2018} and the latter activating in its negative. Feature 356 activates in various rain-forest regions. We show the top 10 features in terms of total summed activation on both timescales in Appendix Fig. \ref{fig:top10}.

To untangle what features have been dynamically inferred by GraphCast and what features may be present in the atmospheric data itself, we repeat the SAE training process on activations taken from a randomly initialized GraphCast. We find that these features also have diurnal and seasonal peaks (Appendix Figure \ref{fig:randomtime}), reflecting underlying variation in the atmospheric data. For a qualitative comparison, we also show the top 20 most highly activating features in SAEs trained on the randomly initialized GraphCast and the trained GraphCast in Appendix Figure \ref{fig:randommag}. The features from the trained GraphCast appear significantly more correlated with atmospheric phenomena, while random features mainly activate on particular regions of the globe.


\subsection{Grid-locked features}

\begin{figure}[ht]
    \centering
    \includegraphics[width=1.0\linewidth,  trim={0cm 0 0cm 0}, clip]{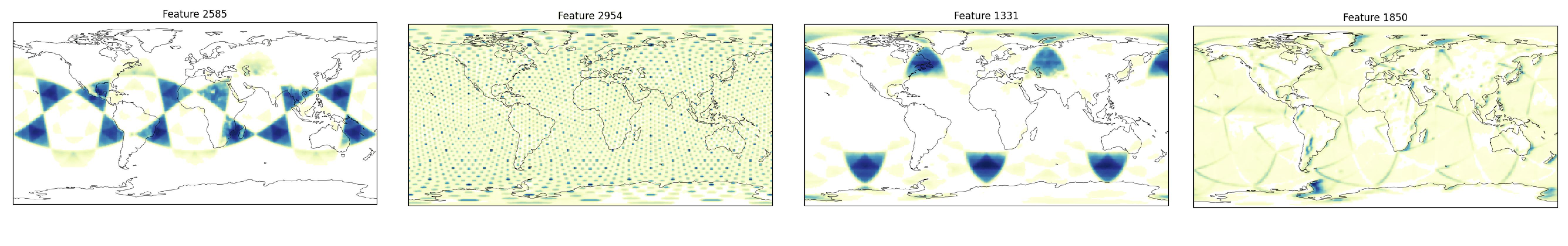}
    \caption{Grid-locked features: spurious features activate on the grid representation of GraphCast.} 
    \label{fig:gridlocked}
\end{figure}

We also note the presence of grid-locked features, that is, those that activate in an average sense on the grid structure of GraphCast (Fig \ref{fig:gridlocked}). From an interpretability perspective, such features are undesirable, as they reflect concepts related to the underlying model architecture as opposed to the physical content of the data. Whether they are also undesirable from a modeling perspective, i.e. in the training of deep-learning atmospheric physics models, is an open question. On the one hand, the multi-scale message passing architecture of GraphCast enforces a natural heterogeneity on the nodal embeddings---some nodes include long-range connections and others do not, so we might expect different representations at some level. On the other, however, significant nodal features that differ from the underlying data may imply a degree of architecture bias where ideally the physical parameterization is invariant to transformations, potentially affecting performance. In either case, we believe that the extraction of such features demonstrates the potential feedback between interpretability approaches and model development.

\subsection{Sparse probing of extreme weather features}

Let us now focus on features concerning particular physical events using a technique known as sparse probing \cite{Gurnee2023FindingProbing, Gao2024ScalingAutoencoders}. In \cite{Kim2025AWeather}, the authors crowdsourced a data-labeling task and curated a large dataset spanning dozens of years of binary masks of TCs, ARs, and atmospheric blocking events with full atmospheric coverage. Focusing on TCs, we take the years 2019, 2020, 2021 (our validation set) and extract full atmospheric masks of TC presence that correspond to the 6 hour time steps on which we have GraphCast activation data (00hr, 06hr, 12hr, and 18hr UTC). We then project the atmospheric labels onto the GraphCast grid using the same geometric relations used to encode atmospheric data onto GraphCast's internal mesh representation (see \cite{Lam2023LearningForecasting} for details) such that we have a label on each node of the GNN's mesh (with non-binary values due to interpolation treated as zeros). Then for each node, depending on our hyperparameter choice, we have $n_l$ features and a single binary value indicating the presence of a TC. Recall that the $k$-sparsity constraint means that only $k$ of these features will be non-zero for a given node. We then train a logistic probe to output a probability of TC mask present on a per-feature basis as

\begin{equation}
    p_{i,j} = \sigma(a \alpha_{i,j} + b)
\end{equation}
with loss
\begin{equation}
    \mathcal{L}_j =
-\frac{1}{2 N_1} \sum_{i:\,y_i=1} \log p_{i,j}
-\frac{1}{2 N_0} \sum_{i:\,y_i=0} \log (1 - p_{i,j}) .
\end{equation}
In words, we have $N_1$ positive samples representing the presence of a TC and $N_0$ negative samples representing the absence. Each probe takes a single feature $j$ on a single node $i$ at layer $l$ and predicts the probability of a positive sample. The loss is balanced to account for the large asymmetry between positive and negative samples ($N_0 \approx 10000 N_1$), even though every time step we sample has a TC present somewhere in the atmosphere. We can also do the same process for single neuron activations (that is, the values of the node embedding without an SAE decomposition).

\begin{figure}[ht]
    \centering
    \includegraphics[width=0.95\linewidth,  trim={0cm 0 0cm 0}, clip]{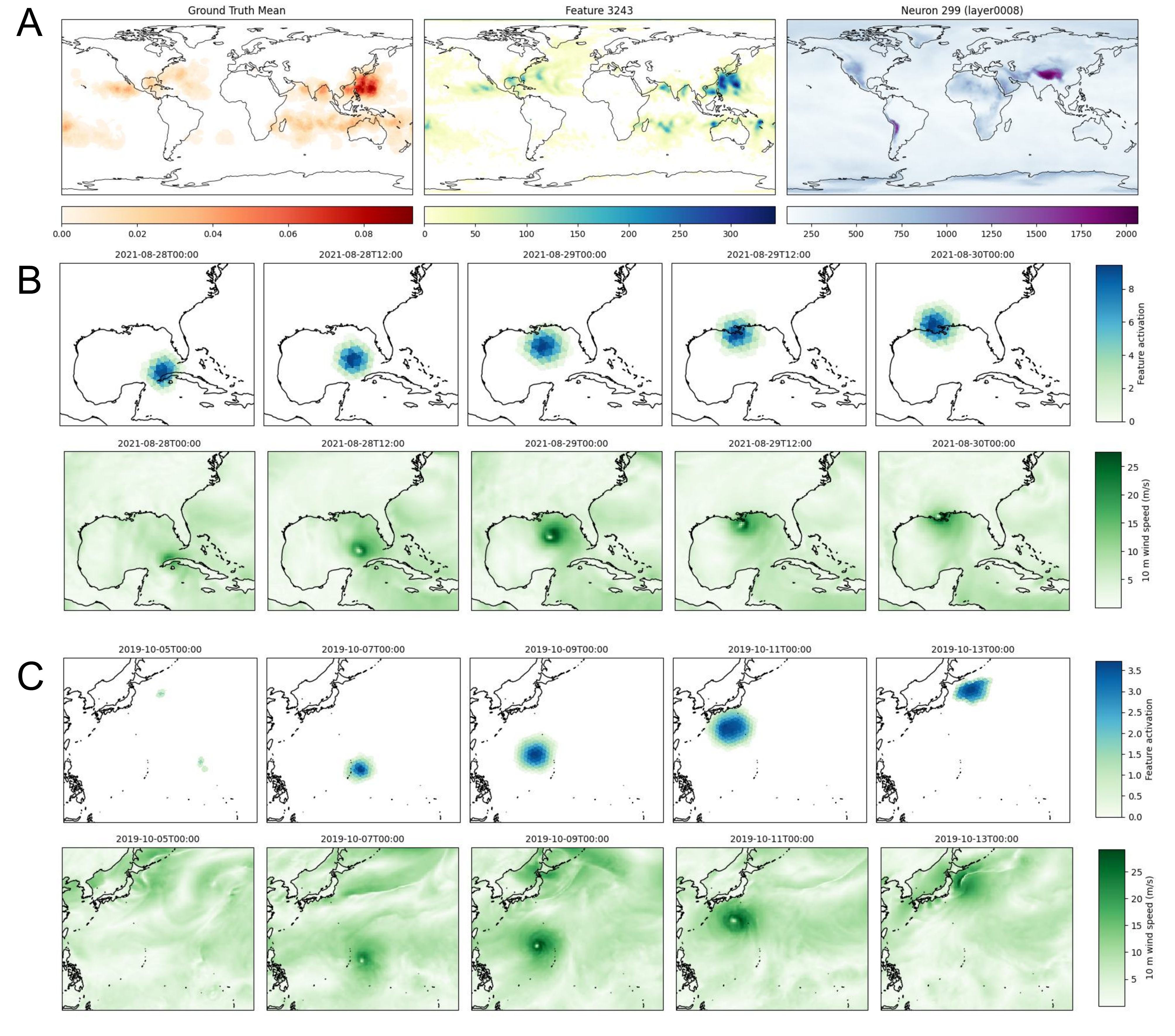}
    \caption{TC feature interpretability. \textbf{(A)} Comparison of three year average of the ground truth TC dataset \cite{Kim2025AWeather}; the highest F1 score feature, feature 3243 from our $l=8$, $k=32$, $n_l=4096$ SAE; and one of the most informative neurons at layer 8, neuron 19. The average activation of feature 3243 aligns strongly with the ground truth dataset, while the single neuron is mostly uninformative. \textbf{(B)} One example of localized feature activation. Top: feature 3243 activation for the duration of Hurricane Ida (2021) as it makes landfall. Bottom: 10 meter wind magnitudes from the same times. \textbf{(C)} A second example of localized feature activation. Top: feature 3243 activation for Typhoon Hagibis (2019) in the Pacific Basin. Bottom: 10 meter wind magnitudes from the same times. } 
    \label{fig:hurricanes} 
\end{figure}

Figure \ref{fig:hurricanes} shows the resulting best single feature extracted from the $n_l=4096$, $k=32$, $l=8$ SAE, whose probe achieves an F1-score of 0.48. When contrasted with the ground-truth dataset, our feature shows remarkable accuracy, activating in the presence of TCs in all major ocean basins (Fig \ref{fig:hurricanes}A). More detailed analysis shows that the feature closely tracks the activity of individual storms (Fig \ref{fig:hurricanes}B, \ref{fig:hurricanes}C), and performs far better than a probe trained only on neurons, which achieves a best F1 score of 0.01 (essentially a randomized predictor). A similar analysis for the most predictive features for atmospheric rivers is carried out in Appendix \ref{sec:rivers} where the main results are summarized in Appendix Fig. \ref{fig:rivers}. We report the best probes for a range of hyperparameter choices for both TCs and ARs in Appendix \ref{sec:hyperparameters}, as well as the mean activation on the HURDAT storm dataset in Appendix Fig. \ref{fig:hurdat}.



\section{Testing internal abstractions}


There are several reasons we might seek a more complete causal account of the features in GraphCast. The first is related to our initial motivation: to probe how complete the model's internal abstraction of a given physical process is, and by extension what exactly it has learned. For instance, it is possible that a feature mostly activating around hurricanes is actually just thresholding vorticity at midlatitudes, and so is only a superficial hurricane detector without encoding a deeper physical abstraction. A true hurricane then would have no disentangled representation in GraphCast, and it would be difficult to give any account of what GraphCast has ‘learned’ about hurricanes. 

Another reason is for model control as it relates to circuit analysis: if by modifying a single feature we can alter model output interpretably, and further by studying which other features modify this feature (which input fields modify these features, and so on) we can begin to construct a mechanistic account of how hurricanes are predicted inside of GraphCast. A necessary precursor for such an explanation is to study whether we can measurably steer outputs with simple modifications of the ‘atoms’ of this circuit. 

\subsection{Feature modification}

We therefore seek to directly analyze a learned abstraction by modifying the feature and measuring the change in the model's output, as a way to directly measure the effect that this abstraction has on the model's predictive power. In a linearized sense, this would be similar to viewing the gradient of a model prediction with respect to a feature. In many contexts, this is an appropriate method of analysis \cite{Marks2025SparseModels}. 
But ultimately causal analysis and in-situ modification go hand in hand \cite{Geiger2025CausalInterpretability}. 


To modify a feature, we first decompose the desired model layer activations into a residual error term due to SAE reconstruction as 

\begin{equation}
    \mathbf{v}_i^{l}=\mathbf{\hat{v}}_i^{l} + (\mathbf{v}_i^{l} - \mathbf{\hat{v}}_i^{l})
\end{equation}
so that the model works as intended without the influence of reconstruction errors on further processing. We then define a modification vector $\mathbf{s}_{f, \gamma}$ that contains the value $1+\gamma$ at index $f$ and 1 in all other entries. For feature $f=3243$, for example, a positive value of $\gamma$ would indicate artificially increasing the activation of the feature, while a negative value would indicate its inhibition. On the forward pass of the model, we compute

\begin{equation}
    \tilde{\alpha}_i^l = \mathbf{s}_{f, \gamma}\: \odot \: \alpha_i^l
\end{equation}

\begin{equation}
    \tilde{\mathbf{v}}_i^l = W_{dec}\tilde{\alpha}_i^l+\mathbf{b}_{pre}
\end{equation}
and set the forward pass activations to be

\begin{equation}
    \mathbf{v}_{i,mod}^{l}=\tilde{\mathbf{v}}_i^{l} + (\mathbf{v}_i^{l} - \mathbf{\hat{v}}_i^{l})
\end{equation}
so that the error term stays unchanged while we force the other main reconstructed component of the residual layer.

\begin{figure}[ht]
    \centering
    \includegraphics[width=1.0\linewidth,  trim={0cm 0cm 0cm 0cm
    }, clip]{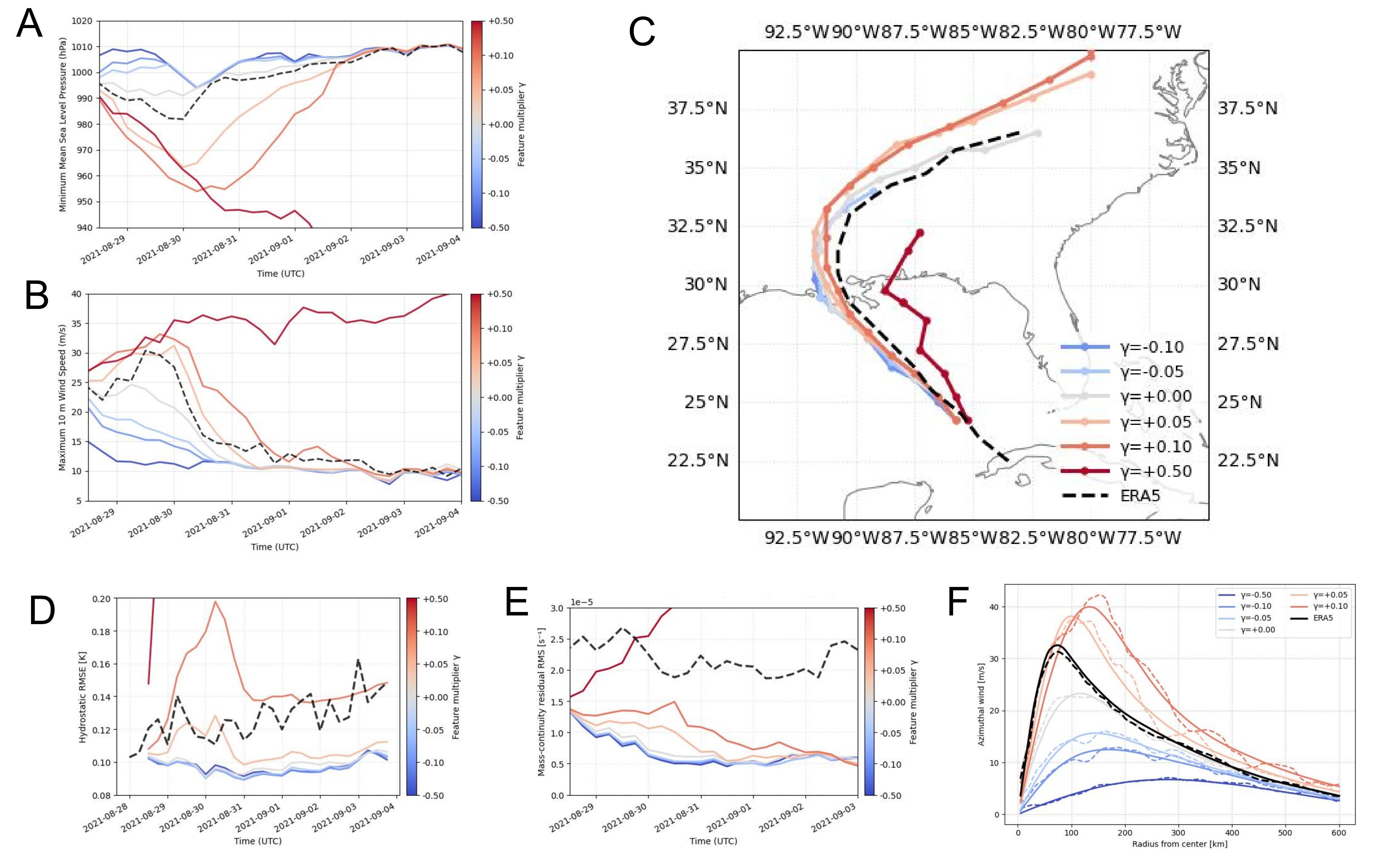}
    \caption{Artificially increasing (decreasing) a hurricane feature (feature 3243) during the GraphCast forward increases (decreases) hurricane strength prediction in a physically plausible manner \textbf{(A,B)} Maximum 10 meter wind speeds and minimum mean sea level pressure for all steered GraphCast hurricane forecasts, along with comparison to ERA5. \textbf{(C)} Paths of modified hurricanes. \textbf{(D,E)} Mass continuity and hydrostatic residual show the range of modified $\gamma$ parameters for which physical consistency is maintained. \textbf{(F)} Gradient wind balance demonstrates that feature modification increases pressure gradient and wind speed in a manner that still maintains force balance around eye of hurricane. Solid lines show measured azimuthal wind speed in a frame relative to the hurricane. Dashed lines show theoretical wind speeds calculated from measured pressure gradients.} 
    \label{fig:steering}
\end{figure}

Figure \ref{fig:steering} shows our results for the modification of feature 3243 at layer 8. We show results for Hurricane Ida, but all tropical cyclones we tested responded in a similar manner. The Hurricane Ida field is initialized at 2021-08-28T12 as it begins to rapidly intensify into a Category 4 storm. Then we take values of $\gamma=[-0.5, -0.1, -0.05, \: 0.0, \: 0.05, \: 0.1, \: 0.5]$, the first three corresponding to an inhibited feature and the last three to an amplified one, and run GraphCast for 40 autoregressive steps (10 days). At each forward pass, at the eighth layer, we modify the single feature with the chosen value of $\gamma$ and continue the forward pass identically in every other sense, repeating at each time step. 

While before we showed that this particular feature correlated with hurricane strength, Figure \ref{fig:steering} (A,B) shows a monotonic trend with $\gamma$ to stronger and weaker hurricane depending on its sign. The largest positive value of $\gamma$ corresponds to a runaway and unphysical response, but all other values follow the typical development of the storm, maintaining the rough path and dissipating as it makes landfall (Figure \ref{fig:steering} (C)). This result constitutes a direct causal relationship between the values of this particular feature and the strength of hurricanes, solidifying this feature as a building block for future circuit analyses.

\subsection{Conservation laws and force balances}

Next, to examine the physical consistency of the internal abstraction, we ask whether the fields produced by modifying the feature adhere to known conservation laws and force balances.

The dynamical core of ERA5 before data assimilation directly enforces hydrostatic balance

\begin{equation}
    \frac{\partial p}{\partial z} = -\rho g,
\end{equation}
where $p$ is the pressure, $z$ is the vertical coordinate, $\rho$ is the mass density, and $g$ is the acceleration due to gravity, which in the discretization scheme along pressure levels can be approximated \cite{Subramaniam2025ImposingPrediction, EuropeanCentreforMedium-RangeWeatherForecasts2024IFSImplementation} as

\begin{equation}
    \frac{T_{v,k}+T_{v,k-1}}{2}=-\frac{g}{R_d\log{(p_k/p_{k-1})}}\left(Z_k-Z_{k-1}\right)
\end{equation}
where $T_v$ is virtual temperature, $Z$ is geopotential height, and $R_d=287$ is the gas constant for dry air. Measuring the residual of this equation indicates how far out of hydrostatic balance a given field is, and can act as a diagnostic for unphysicality. As an important caveat, ERA5 fields themselves are not in perfect hydrostatic balance due to its data assimilation process, and can deviate by order 0.1 K at 500 hPa \cite{Subramaniam2025ImposingPrediction}. As hydrostatic balance is essentially a force balance that neglects the effects of vertical accelerations, there is reason to believe that especially in the vicinity of a hurricane it would be violated.

A complementary diagnostic is mass conservation. This law enforces the constraint that a parcel of air cannot gain or lose mass, and on ERA5 pressure coordinates takes the form \cite{Malardel2019DryIFS}:

\begin{equation}
    \nabla_h\cdot\mathbf{v_h} + \frac{\partial \omega}{\partial p} = 0 ,
\end{equation}
which states that the horizontal gradient of the horizontal wind must balance the vertical gradient (in pressure coordinates) of the vertical velocity $\omega$. 

We expect these two laws to hold approximately everywhere in the atmosphere. We find that they largely do in integrated root-mean-square error (RMSE) for the evolution of Hurricane Ida, offering evidence for the physical consistency of the result fields (Figure \ref{fig:steering}D,E). But there are also coarse force-balances that one expects to hold around a tropical cyclone specifically. In particular, in a cylindrical coordinate frame centered at the eye of a hurricane, the centrifugal force of the azimuthal velocity $v_{\theta}^2/r$ and the Coriolis force $fv_{\theta}$ is approximately balanced by the pressure gradient force $\frac{dZ}{dr}$ \cite{Sun2025CanCyclones, H.E.Willoughby1990GradientCyclones}:

\begin{equation}
  g\frac{\partial Z}{\partial r} = \frac{v_{\theta}^2}{r} + fv_{\theta}  ,
\end{equation}
which allows us to produce a reconstructed wind field using just the pressure gradient and compare it to the measured azimuthal wind field. Figure \ref{fig:steering}F shows that for intermediate feature modification values, this force balance is strongly maintained, demonstrating that modifying the value of the feature increases wind speed and pressure in the \textit{unique} correspondence that allows the final resulting field to remain physical. Whether this is a robust probe of the model's internal concept or more a measure of the model's ability to self repair \cite{McGrath2023TheComputations} will be an object of future research. In either case, as the predictions generated by this feature modification process are largely physical, we note that in addition to testing a model's internal TC abstractions, this process could be used to generate unlikely scenarios of storm evolutions without resorting to expensive ensemble approaches.

\section{Conclusion}

We have shown that sparse autoencoders applied to the internal activations of the large data-driven physics model GraphCast extract highly interpretable features of GraphCast's internal processing and the underlying dynamical system. We demonstrated a probe of these abstractions by precisely modifying a tropical cyclone feature and observing the correct and physics-constrained response of the underlying model prediction.

By studying the internal representations inside data-driven physics models like GraphCast, we hope that greater model trust may be achieved. We also suggest that these methods may be a profitable direction for studying the underlying dynamical system itself. As we are able to extract interpretable structures from the neurons of a model trained purely to predict future states of data, with no interpretability constraints, perhaps we no longer have to pursue a two-pronged approach of, on the one hand, small, interpretable methods, which have considerable scientific value but cannot make practical predictions, and, on the other hand, large predictive methods that have operational value but are less valuable for generating scientific insight. 


GraphCast and other data-driven models must have leveraged some reduced-order model to parameterize the time-stepping of their respective dynamical system, as evidenced by their accuracy and time cost. As we have shown that in some cases this reduced-order model can be cast in an interpretable neuron basis, it is possible that through analysis of these features and the way they interact, truly novel physical pathways could be discovered---namely the pathways that have allowed data-driven models to achieve their empirical success. We suggest that a future research direction that may find interesting answers to these questions would be to build mechanistic accounts for the interaction of features leading to model predictions, or to train specifically sparse models \cite{Gao2025Weight-sparseCircuits} to have interpretable computations.

\section*{Acknowledgments}
Some of the computing for this work was performed on the Sherlock cluster at Stanford University. We thank Stanford University and the Center for Computation at the Stanford Doerr School of Sustainability for providing the computational resources and support that were essential to our research outcomes. We thank Cristobal Eyzaguirre, Robert King, and Elias Huseby for fruitful discussions over the course of the study. We thank Elana Simon for suggesting a comparison of trained SAEs to a random baseline.

\paragraph*{Funding:}
T.M. acknowledges financial support from a Stanford Graduate Fellowship and from the National Science Foundation Graduate Research Fellowship Program under Grant No.~DGE-1656518.



\bibliographystyle{unsrtnat}
\bibliography{references,references-2}

\newpage
\appendix

\section{GraphCast architecture}
\label{sec:graphcast}

We refer readers to \cite{Lam2023LearningForecasting} for a complete description of GraphCast's architecture. Here, we give a high-level overview. GraphCast is a graph neural network (GNN) operating in an encode-process-decode framework, trained to predict future states of the atmosphere given past and present data as

\begin{equation}
    X^{t+1} = \textrm{Graphcast}(X^t, X^{t-1}).
\end{equation}
First, the state of the atmosphere is encoded onto the internal mesh representation of Graphcast using a trainable graph encoder to obtain node embeddings $\mathbf{v}^0$ and edge embeddings $\mathbf{e}^0$. Next, 16 layers of message passing on the internal mesh are carried out to process the internal state. At each layer $l$ of message passing in GraphCast, the $n_n$ nodes and $n_e$ edges update their embeddings, $\textbf{v}^l\in \mathbb{R}^{n_d\times n_n}$ and $\textbf{e}^l\in \mathbb{R}^{n_d\times n_e}$, via message-passing mediated by multi-layer perceptrons. When message passing is complete, the embeddings are projected back onto the atmospheric grid using a trainable graph decoder. We focus on the internal message-passing step, which can be written as an update rule on the node embeddings as

\begin{equation}
    \mathbf{v}^{l+1}_i = \mathbf{v}_i^{l} + \textrm{MLP}^{l+1}_{node}\left(\mathbf{v}_i^l, \: \: \sum_{e^{l+1}_{v_s\rightarrow v_r}:v_r=v_i}\mathrm{MLP}^{l+1}_{edge}(\mathbf{e}^l_{v_s\rightarrow v_r},\mathbf{v}^l_s, \mathbf{v}^l_r)\right),
\end{equation}
where $\mathbf{v}_i^l\in \mathbb{R}^{n_d}$ is the embedding of node $i$ at layer $l$ and $\mathbf{e}^{l}_{v_s\rightarrow v_r}$ is the embedding of the edge between sender node $s$ and receiver node $r$ at layer $l$. The parameter $n_d$ is the latent dimension of both node and edge embeddings, and for GraphCast is set to $n_d=512$.

The mesh of GraphCast is generated by iteratively refining an initial icosahedron by dividing each face into four smaller faces. The nodes at each level are a subset of the higher level refined mesh, and all edges are included from each refinement step to allow for natural long-range connection. For the high resolution version that we analyze, six mesh refinement steps result in a total of $n_n=40,962$ nodes and $n_e=327,660$ edges that are fixed through each round of message passing. This presents a challenge for our interpretability approach, as nodes included at earlier stages of refinement will have longer-range and more numerous connection, whereas all nodes are treated as equivalent in our SAE training. As we demonstrate in Fig \ref{fig:gridlocked}, this refinement process manifests in the learned features of GraphCast.

We focus directly on the node embeddings, although we note that a more complete account of GraphCast would also include the edge embeddings, which can be thought of as a kind of attention mask. Indeed, later versions of the graph-based architecture of Graphcast explicitly utilize attention networks for the message passing step \cite{Alet2025SkillfulMarginals}. From a data perspective, capturing edge embeddings in GraphCast requires an order of magnitude more data storage capacity. Further, the node embeddings serve as a closer analogue of the residual stream in transformer models, although information is passed between layers in the edge embeddings as well as the node embeddings.

\section{Relation to lower-dimensional data-driven physics models}
\label{sec:sindy}

A comparison to dynamical systems may be interesting for some readers. For instance, consider the goal of SINDy \cite{Brunton2016DiscoveringSystems}, a method that seeks to parameterize a time operator as a sparse linear combination of a pre-defined dictionary of candidate functions, or $\dot{x} = \sum \alpha_i f_i(x)$. Here, interpretability is enforced by construction of the dictionary vectors, which are simple functions of the input---for instance $\cos(x)$, $x^2$, $\exp{x}$, etc. The assumption here is not unrelated to the linear representation hypothesis. To work, SINDy posits that there is a more interpretable basis that captures the majority of the variation of the data in question. In an SAE, we are concerned instead with the space of neurons and want to find the corresponding basis in an unsupervised fashion, with a similar emphasis on sparsity. Since GraphCast runs quickly and accurately, we might expect that it has already learned some simple representation, and all we have to do is extract it. This approach is somewhat contrary to work that has focused on data-driven techniques that employ deep multi-layer perceptrons (MLPs) but that enforce interpretability at some intermediate layer; notable examples are \cite{Lusch2018DeepDynamics} and \cite{Champion2019Data-drivenEquations}. Here, the fundamental task is not to perform a better prediction than traditional methods, but to find an interpretable transformation of the dynamics. The success of our general approach would be to show that interpretable representations emerge spontaneously when a physics model performs sufficiently well on a prediction task, but such a general claim is not fully justified by our single model analysis. Rather, we show that it \textit{can} happen in principle and \textit{does} in GraphCast.

\section{Training k-sparse autoencoders} 
\label{sec:sae}
$k$-sparse autoencoders enforce $k$-sparsity by construction, in contrast to $L_1$ norm penalization \cite{bricken2023monosemanticity} that seeks to optimize the $L_0$ norm indirectly. We carried out preliminary experiments with the $L_1$ autoencoder architecture, but found more difficulty controlling the presence of ``dead'' features---a phenomenon where a small set of features activates for every example, considerably under-utilizing the full $n_l$-dimensional feature space. We were more successful applying the mitigation strategies of \cite{Gao2024ScalingAutoencoders}, and the direct $k$-sparsity allows us to store only $k$ feature activation per node embedding instead of the full $n_l$-sized activation vector, considerably reducing our storage requirements.

We begin by fixing $l=8$, a middle layer, and trained SAEs at a range of sparsities $k=16, \: 32, \: 64, \: 128, \: 256$ and latent dimensions $n_l=2048, \: 4096, \: 8192, \: 16384$. To reduce the number of dead features accumulated during training, we use an auxiliary loss as in \cite{Gao2024ScalingAutoencoders}. We define a dead latent by an index of $\boldsymbol{\alpha}$ that has not appeared in the top $k$ features for the past 5,000,000 node embeddings. The activation function $\mathrm{TopKAux}(\mathbf{x})$ is defined similarly to TopK, but instead of zeroing out all but the top $k$ values of $\mathbf{x}$, it zeros out all but the top $k_{aux}$ dead features. We set $k_{aux}=512$ for all training runs. We then use these features to reconstruct the residual error of the standard pass of the SAE, penalizing the network for under-utilizing the latent space: 

\begin{equation}
    \mathbf{\tilde{v}}_i^l = W_{dec}\mathrm{TopKAux}(W_{enc}(\mathbf{v}_i^l-\mathbf{b}_{pre}))+\mathbf{b}_{pre} .
\end{equation}
Our loss function is a combination of MSE on the node embeddings and this extra penalty:

\begin{equation}
    \mathcal{L} = ||\mathbf{v}_i^l-\mathbf{\hat{v}}_i^l||_2+\beta ||(\mathbf{v}_i^l-\mathbf{\hat{v}}_i^l)-\mathbf{\tilde{v}}_i^l||_2 ,
\end{equation}
where $\beta=\frac{1}{32}$ is a hyperparameter, which we set to the value used in \cite{Gao2024ScalingAutoencoders}.

\begin{figure}[ht]
    \centering
    \includegraphics[width=0.95\linewidth]{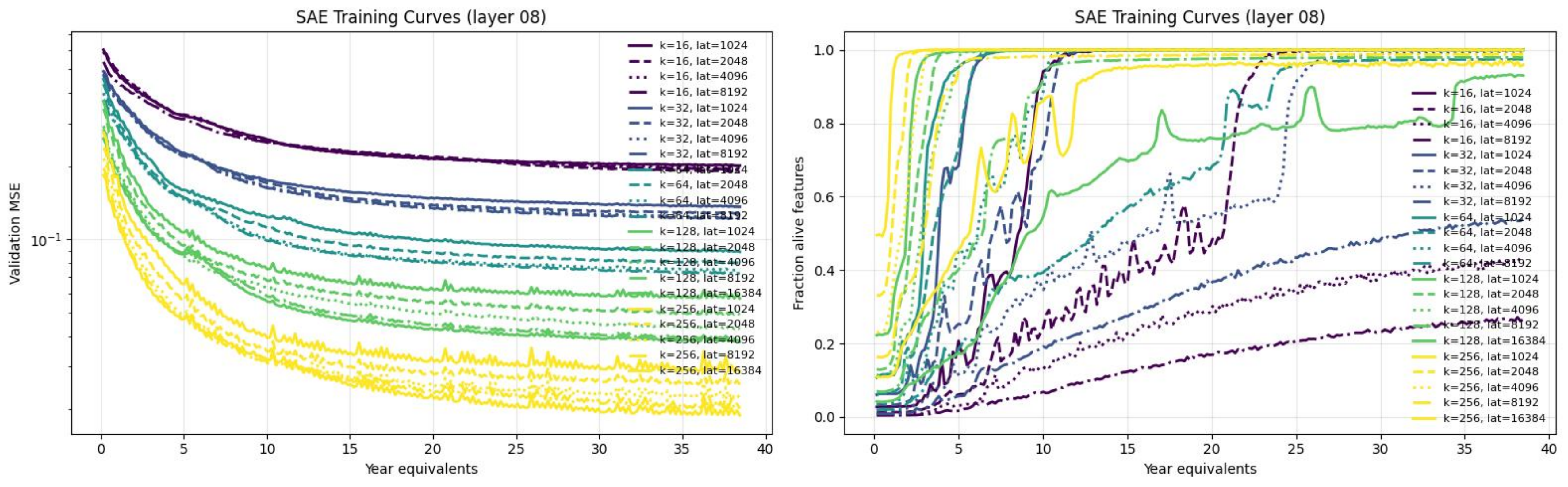}
    \caption{(Left) Training loss as a function of year equivalents of node embeddings. (Right) Fraction of alive features.}
    \label{fig:training_loss}
\end{figure}

In all cases we used a batch size of 8192 and a fixed learning rate $l=3\times 10^{-4}$. Figure \ref{fig:training_loss}A shows the mean-squared validation error as a function of the input data. To reach the loss plateau, at least 20 years of input data was required, but it is possible that this could be adjusted by modifying the learning rate, among other hyperparameters. Figure \ref{fig:training_loss}B
shows the fraction of alive features during the training process. Note that sparser autoencoders (lower $k$) take a significantly longer time to ameliorate their dead features, especially when there are many of them (larger $n_l$). For some of the sparser autoencoders, the number of features never reached full saturation. Since we chose not to recycle any data, other design choices like the values of $k_{aux}$ or $\beta$ may be helpful in training these ultra-sparse models.

\begin{figure}[ht]
    \centering
    \includegraphics[width=0.95\linewidth]{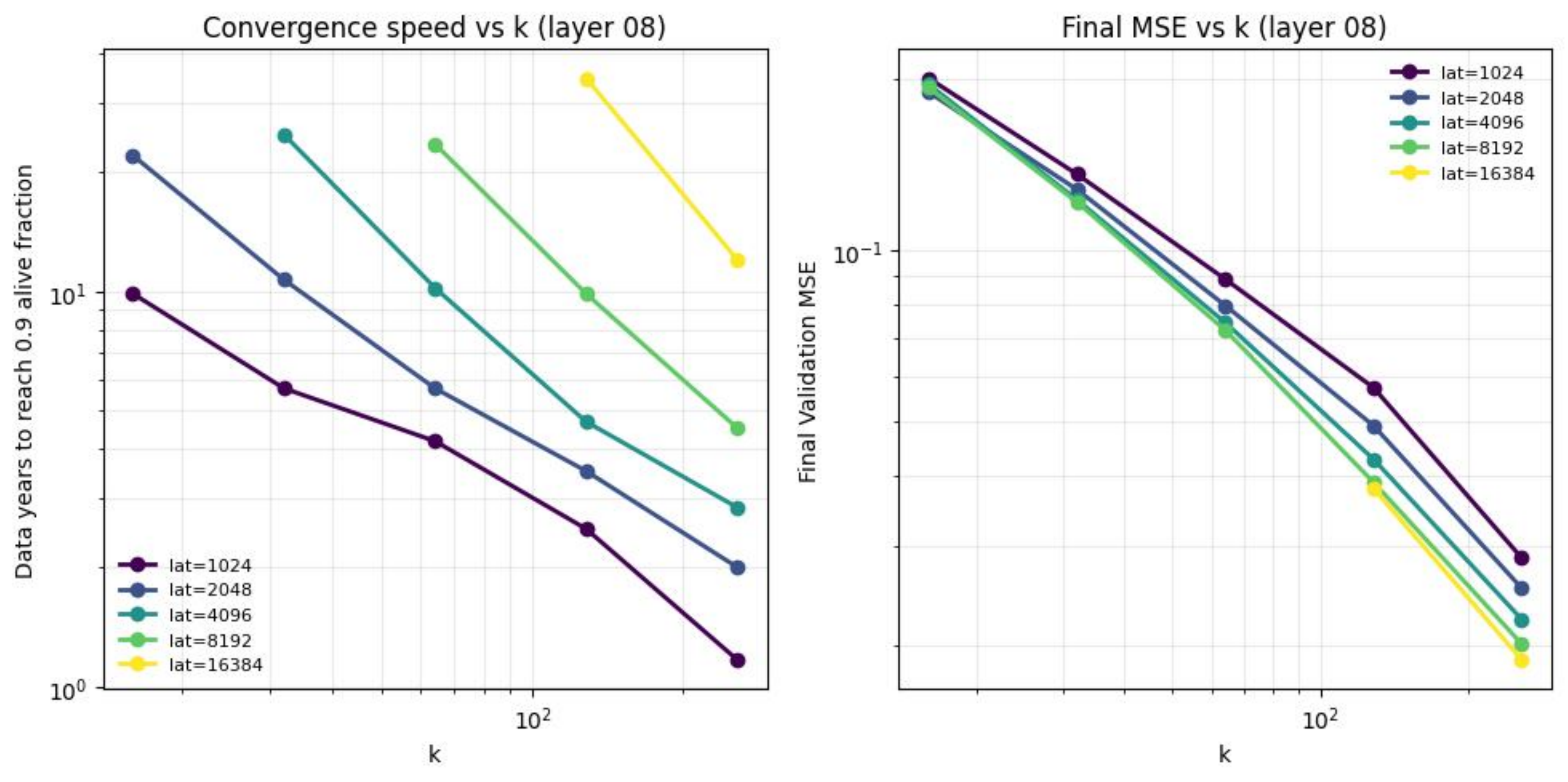}
    \caption{(Left) Amount of data needed to reduce number of dead features across model hyperparameters (Right) Final validation loss across model hyperparameters.}
    \label{fig:power_laws}
\end{figure}

We plot the final loss attained by each run as well as the number of node embeddings to reach full feature saturation as a function of our input hyperparameters $k$ and $n_l$ in Figure \ref{fig:power_laws}. For each value of $n_l$, the mean-squared error loss follows an approximate power law in $k$. For the more sparse models, the latent dimension does not have a measurable effect on validation loss, but this could also because these models are unable to rectify their dead features during training. These curves represent a Pareto-front of sparsity versus reconstruction error, and would be instructive if comparing different architectures (e.g., $L_1$-norm sparse autoencoder). One advantage of utilizing the $k$-sparse architecture is that the $L_0$ norm is directly fixed and does not have to be indirectly controlled via $L_1$-penalization or otherwise. We also note that the amount of data required to reach 90\% features utilized is approximately a power law with different offsets depending on $n_l$. A data point is left out of the plot if the training run never reaches the 90\% alive threshold.

\section{Atmospheric river sparse probing}
\label{sec:rivers}

\begin{figure}[ht]
    \centering
    \includegraphics[width=0.95\linewidth,  trim={0cm 0 0cm 0}, clip]{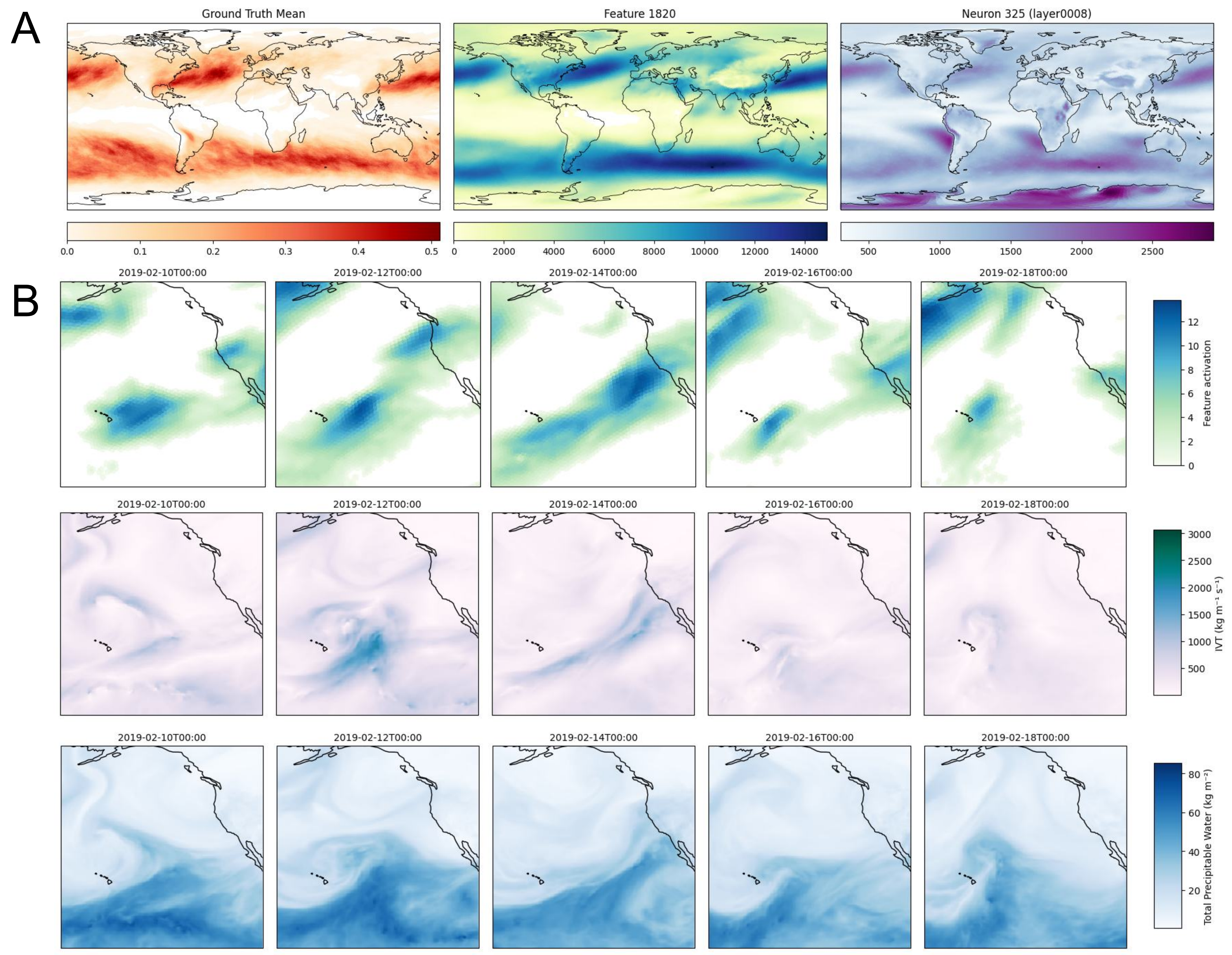}
    \caption{AR feature interpretability. \textbf{(A)} Comparison of three year average of the ground truth AR dataset \cite{Kim2025AWeather}; the second highest F1 score feature, feature 1820 from our $l=8$, $k=32$, $n_l=4096$ SAE; and one of the most informative neurons at layer 8, neuron 325. The average activation of both feature 1820 and the most informative neuron align strongly with the ground truth dataset. \textbf{(B)} One example of localized feature activation. Top: feature 1820 activation for a significant Pineapple Express event in February 2019. Middle: IVT fields, the typical measure of AR activity. Bottom: TPW fields, related to but not equivalent to water vapor transport.} 
    \label{fig:rivers} 
\end{figure}

To complement the TC analysis in the main text, here we visualize the results of a strongly predicting feature on AR detection. We visualize the second best feature (feature 1820) instead of the best (feature 1006), as the highest F1-score features activated with considerably more sparsity. The results are shown in Figure \ref{fig:rivers}. We show a single event corresponding to the Pineapple Express, a well-known atmospheric river affecting the western coast the United States. As shown in the figure, although the AR feature attains lower F1-scores than the best TC features, the AR feature activates strongly in the present of integrated vapor transport (IVT), the strongest signal of atmospheric rivers. We also show total precipitable water to show that the feature does not simply activate in the presence of moisture in the atmosphere.

\section{Probes for different hyperparameters}
\label{sec:hyperparameters}
\begin{figure}[ht]
    \centering
    \begin{subfigure}[b]{0.45\textwidth}
        \centering
        \includegraphics[width=\textwidth]{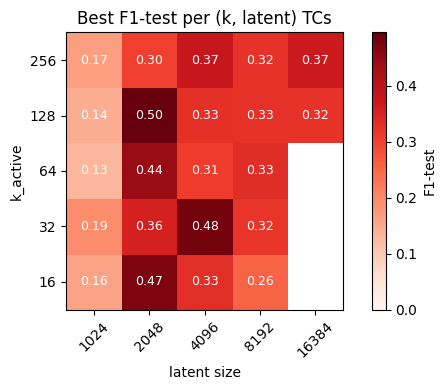}
        \label{fig:plot1}
    \end{subfigure}
    \hfill
    \begin{subfigure}[b]{0.45\textwidth}
        \centering
        \includegraphics[width=\textwidth]{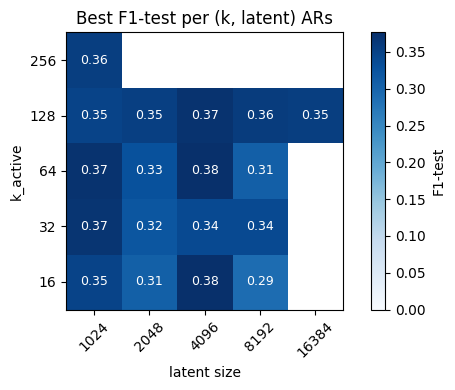}
        \label{fig:plot2}
    \end{subfigure}

    \caption{(Left) F1 scores for single feature logistic probes trained to predict tropical cyclone presence. (Right) Same setup but for predicting atmospheric river presence.}
    \label{fig:f1_scores}
\end{figure}

We trained logistic probes for a range of hyperparameter on TCs and ARs to complement our efficient frontier analysis (Figure \ref{fig:f1_scores}). We find that an intermediate-sized latent layer ($n_l$=2048, 4096) seems optimal for capturing TC presence. We hypothesize this is due to the phenomenon of feature splitting, where one complex feature is broken down into several different features as the dictionary size grows. So it is possible that the larger SAEs learn different features for TCs in different regions of the atmosphere (i.e., different features for hurricanes, typhoons, etc.), violating our broad global categorization and hindering probe performance. For the case of ARs, we found roughly equal representation across all hyperparameter choices, with no clear trend. Interestingly, in contrast to the TC prediction task, a probe trained on just the neuron activations of GraphCast attains a competitive F1-score of 0.30 (worse than all but one SAE, but still strong relative to the best performing models), perhaps due to the strong correlation of ARs with vapor content in the atmosphere. Across all hyperparameter choices, we failed to find features that correspond to atmospheric blocking events (best F1 score of 0.1), the third process included in the \cite{Kim2025AWeather} dataset.

\begin{figure}[ht]
    \centering
    \includegraphics[width=0.95\linewidth]{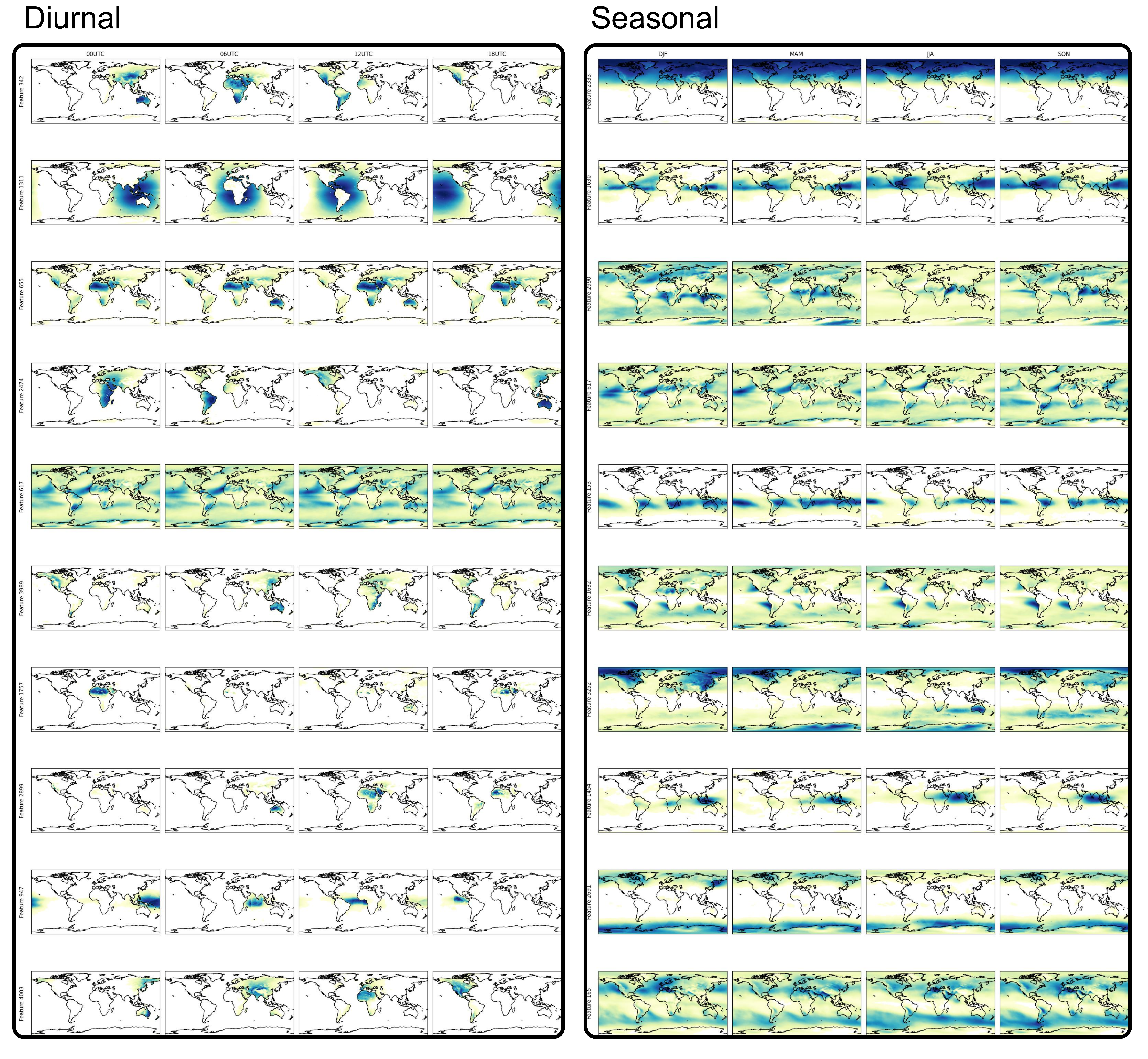}
    \caption{Top 10 energy containing features on diurnal and season timescales.}
    \label{fig:top10}
\end{figure}

\begin{figure}[ht]
    \centering
    \includegraphics[width=0.5\linewidth]{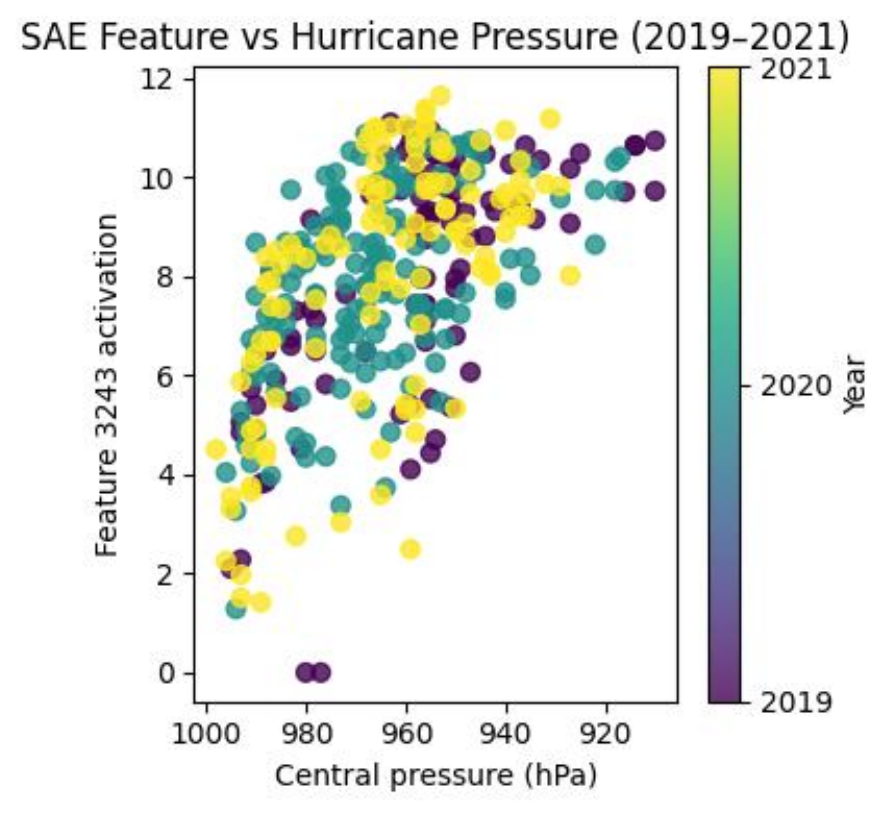}
    \caption{Hurricane feature response against storm central sea level pressure for HURDAT storms across 2019-2021.}
    \label{fig:hurdat}
\end{figure}

\begin{figure}[ht]
    \centering
    \includegraphics[width=0.9\linewidth]{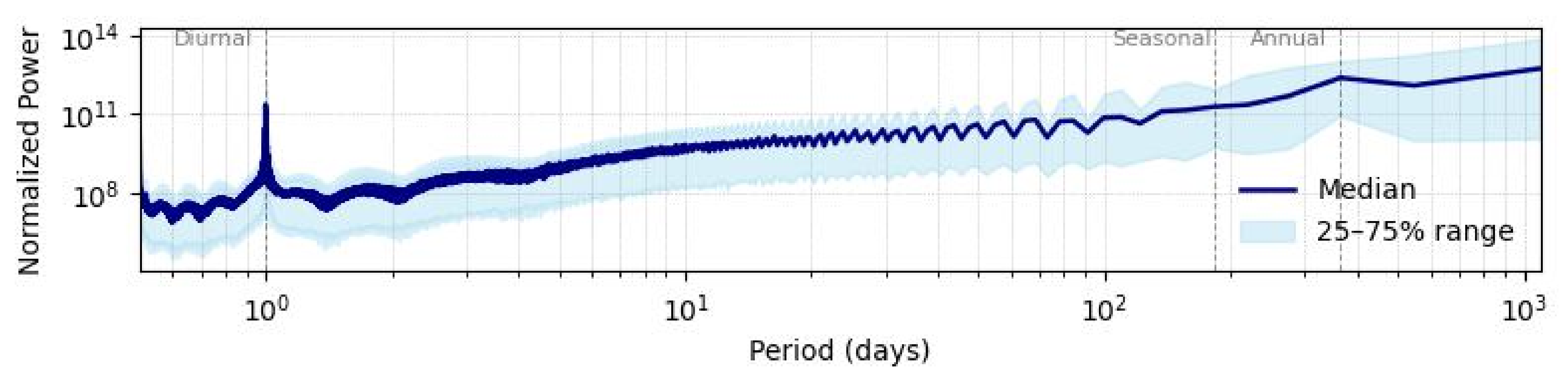}
    \caption{Average power spectrum for SAE trained on randomly initialized Graphcast. Note that there are still peaks at diurnal and seasonal timescales, reflected underlying periodicity in the data.}
    \label{fig:randomtime}
\end{figure}

\begin{figure}[ht]
    \centering
    \includegraphics[width=0.95\linewidth]{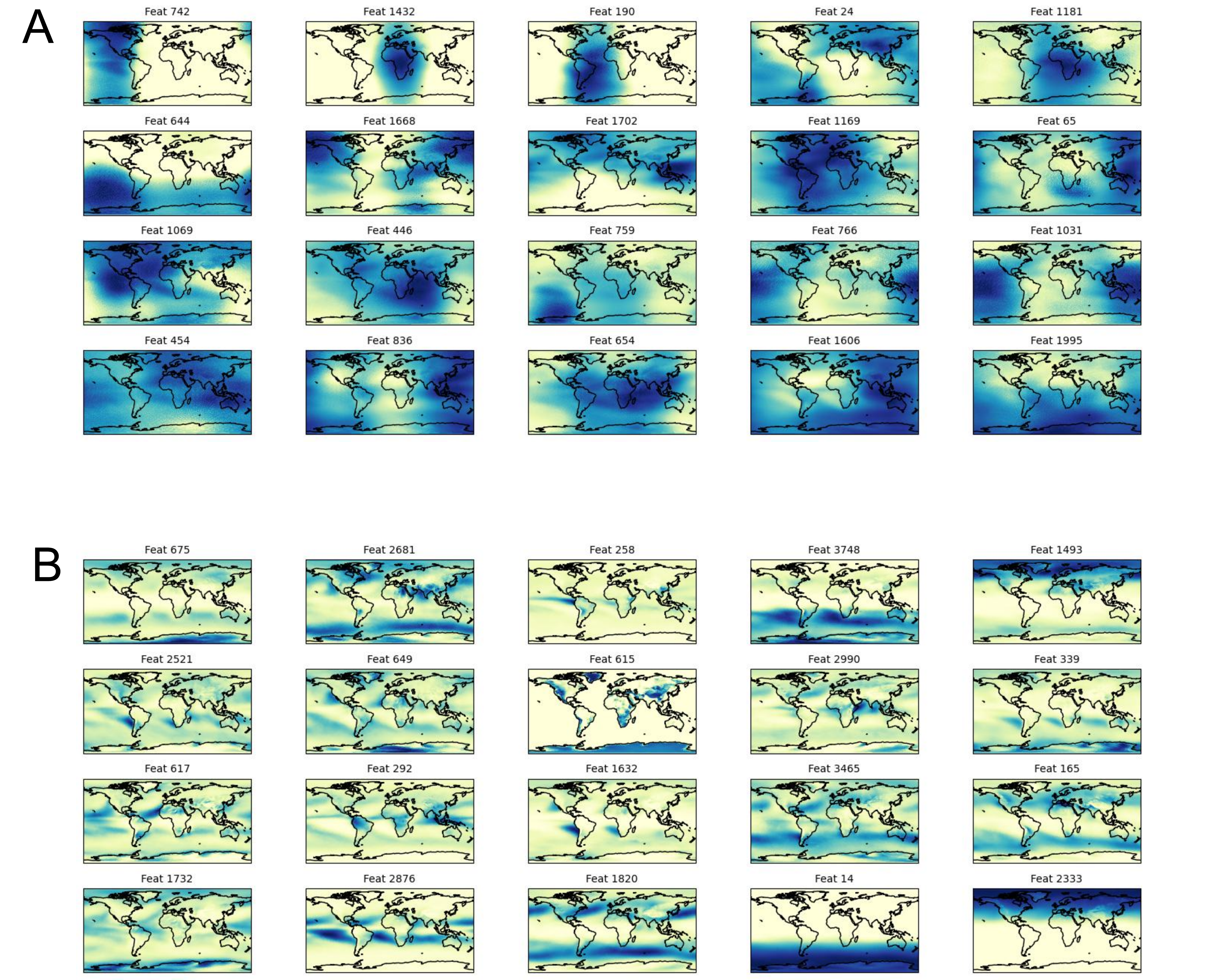}
    \caption{\textbf{(A)} Top 20 largest activating features for SAE trained on randomly initialized GraphCast. Darker colors signify a larger activation. \textbf{(B)} Top 20 largest activating features for SAE trained on trained GraphCast.}
    \label{fig:randommag}
\end{figure}

\end{document}